\documentclass{article}
\usepackage{graphicx,amssymb}
\newcommand{\ket}[1]{\vert #1 \rangle}

\begin{document}
%%%%%%%%%%%%%%%%%%%%%%%%%%%%%%
\title{\bf Degradation of continuous variable entanglement in a
phase-sensitive environment}
\author{Andrea R. Rossi, Stefano Olivares, and Matteo G. A. Paris \\ 
INFM and Dipartimento di Fisica, Universit\`a di Milano, Italia.}
\date{}\maketitle
%%%%%%%%%%%%%%%%%%%%%%%%%
\begin{abstract}
We address the propagation of twin-beam of radiation through
Gaussian phase-sensitive channels, {\em i.e.} noisy channels with 
{\em squeezed} fluctuations. We find that squeezing the environment always 
reduces the survival time of entanglement in comparison to the case of 
simple dissipation and thermal noise. We also show that the survival time 
is further reduced if the squeezing phase of the fluctuations is 
different from the twin-beam phase.
\end{abstract}
%%%%%%%%%%%%%%%%%%%%%%%%%
\section{Introduction}\label{s:intro}
The use of phase-sensitive environments has been addressed by many authors 
\cite{kim,un,du,tri} for preservation of macroscopic quantum coherence. 
In fact, by adding {\em squeezed} fluctuations to dissipation, superpositions 
of states preserve their coherence longer than in presence of dissipation 
alone. A question arises whether or not phase-sensitive environments may be
also used to preserve entanglement. In this paper, in particular, we study 
the behavior of a twin-beam state of radiation (TWB) propagating through a 
Gaussian noisy phase-sensitive channel, in order to investigate whether or not 
squeezing the environment is useful to preserve continuous variable entanglement. 
The answer turns out to be negative. The survival time of entanglement in a 
squeezed bath is always smaller than in purely dissipative or thermal ones, 
and the degradation is more pronounced as far as the bath fluctuations are 
out of the TWB phase. 
%%%%%%%%%%%%%%%%%%%%%%%%%%%%%%%%%%%%%%%%%%%%%%%%%%%%%%%%%%%%%%%%%%%%%%%
\section{TWB in a Gaussian bath}\label{s:interaction}
The propagation of a TWB interacting with a general Gaussian environment, 
can be modeled as the coupling of each part of the state with a
non-zero temperature squeezed reservoir. The dynamics can be
described by the two-mode Master equation 
\begin{eqnarray}
\frac{d\rho_t}{dt} &=& \{ \Gamma (1+N) L[a]+\Gamma (1+N)
L[b]+\Gamma N L[a^{\dag}] + \Gamma N L[b^{\dag}] \nonumber\\
&\mbox{}& \hspace{0.4cm}+ \Gamma M \mathcal{M}[a^{\dag}] + \Gamma
M^* \mathcal{M}[a] + \Gamma M \mathcal{M}[b^{\dag}] + \Gamma M^*
\mathcal{M}[b] \}\rho_t \,, \label{master:squeezed}\;
\end{eqnarray}
where $\rho_t\equiv\rho(t)$ is the system's density matrix at the
time $t$, $\Gamma$ is the damping rate and $N$ and $M$ are the
effective photons number and the squeezing parameter of the bath
respectively (which are assumed to be equal for the two channels). 
$L[O]$ is the Lindblad superoperator, $L[O]\rho_t=O\rho_t O^{\dag}
-\frac12 O^{\dag}O\rho_t - \frac12 \rho_t O^{\dag} O$, and 
$\mathcal{M}[O]\rho_t = O\rho_t O -
\frac12 O O \rho_t - \frac12 \rho_t O O $. Of course, the dynamics
of the two modes are independent on each other.
\par 
Using the differential representation of the superoperators in
equation (\ref{master:squeezed}), the corresponding Fokker-Planck
equation for the two-mode Wigner function $W\equiv W(x_1,y_1;
x_2,y_2)$ is given by \cite{andre:sq}
\begin{eqnarray}
\partial_{\tau} W = \left\{ -\sum_{j=1}^4 \partial_{x_{j}}\:
a_{j}(\underline{x}) + \frac{1}{2} \sum_{i,j=1}^{4}
\partial^{2}_{ x_{i} x_{j}}\: d_{ij} \right\} W\,,
\label{wigner:master:squeezed}
\end{eqnarray}
where, for the sake of simplicity, we put $\underline{x} =
(x_{1},y_{1};x_{2},y_{2}) \equiv (x_{1},x_{2};x_{3},x_{4})$, $\tau
= \Gamma t/\gamma$ and $\gamma = (2N+1)^{-1}$. In equation
(\ref{wigner:master:squeezed}) $a_j(\underline{x})$ and $d_{ij}$
are the matrix elements of the drift and diffusion matrices
$\mathbf{A}(\underline{x})$ and $\mathbf{D}$ respectively, which
are given by
\begin{eqnarray}
\mathbf{A}(\underline{x}) = -\frac12\, \gamma \, \underline{x}\,,
\end{eqnarray}
\begin{eqnarray}
\mathbf{D}= \frac12 \left( \begin{array}{cccc}
\frac{1}{2}+\gamma \, \Re{\rm e}[M] & \gamma \, \Im{\rm
m}[M] &
0 & 0 \\
\gamma \, \Im{\rm m}[M] & \frac{1}{2}-\gamma \, \Re{\rm
e}[M] &
0 & 0 \\
0 & 0 & \frac{1}{2}+\gamma \, \Re{\rm e}[M] & \gamma
 \, \Im{\rm m}[M] \\
0 & 0 & \gamma \, \Im{\rm m}[M] & \frac{1}{2}-\gamma \,
\Re{\rm e}[M]
\end{array} \right)\,.
\end{eqnarray}
Notice that the drift term is linear in $\underline{x}$ 
and the diffusion matrix does not depend on $\underline{x}$. The
positivity of $\mathbf{D}$ requires that $|M|<(2N+1)/2$. Moreover, 
to ensure positivity of the density matrix $\rho_t \geq 0$ we need 
$|M|^2 \leq N(N+1)$, which includes the former condition.
\par
The solution of the Fokker-Planck (\ref{wigner:master:squeezed})
can be calculated analytically. For the case $\Im{\rm
m}[M] = 0$, and considering (without loss of generality) TWB with 
real parameter as the initial state, {\em i.e.} $\rho_0\equiv
\rho_{\rm TWB} =|{\rm TWB}\rangle \rangle\langle\langle 
{\rm TWB} |$, where $|{\rm TWB} \rangle \rangle = \sqrt{1-\xi^2} 
\sum_p\: \xi^{p}\: \ket{p}\ket{p}$, $\xi \in {\mathbb R}$, the 
solution assumes the simple form
\cite{andre:sq}
\begin{eqnarray}
W_{\tau}(x_{1},y_{1},x_{2},y_{2}) = \frac{
\exp\left\{
\displaystyle{
-\frac{(x_{1}+x_{2})^{2}}{4
\Sigma_{1}^{2}}-\frac{(y_{1}+y_{2})^{2}}{4 \Sigma_{2}^{2}}
-\frac{(x_{1}-x_{2})^{2}}{4
\Sigma_{3}^{2}}-\frac{(y_{1}-y_{2})^{2}}{4 \Sigma_{4}^{2}}
}
\right\}
}
{(2 \pi)^2\:
\Sigma_{1}\: \Sigma_{2}\: \Sigma_{3}\: \Sigma_{4}}
\label{wigner:evol:sq}
\end{eqnarray}
where $\Sigma_{j}^2 = \Sigma_{j}^2(r, \Gamma, n_{\rm th}, n_{\rm
s})$, $j=1,2,3,4$, are
\begin{equation}
\begin{array}{lll}
\Sigma_{1}^{2}=\sigma_{+}^{2} e^{-\Gamma t}+D_{+}^{2}(t)\,,
&\quad&
\Sigma_{2}^{2}=\sigma_{-}^{2} e^{-\Gamma t}+D_{-}^{2}(t)\,, \\
\mbox{}\\
\Sigma_{3}^{2}=\sigma_{-}^{2} e^{-\Gamma t}+D_{+}^{2}(t)\,,
&\quad&
\Sigma_{4}^{2}=\sigma_{+}^{2} e^{-\Gamma t}+D_{-}^{2}(t)\,,
\end{array}\label{sigma:1234}
\end{equation}
with $\sigma_{\pm}^2 = \frac14 e^{\pm 2 \lambda}$, $\xi = \tanh \lambda$, and
\begin{eqnarray}\label{sq:D:pm}
D_{\pm}^{2}(t) = \frac{1 + 2 N \pm 2 M}{4} \left(1 -e^{-\Gamma
t}\right)\,.
\end{eqnarray}
For the general case ($M$ complex) the analytical solution of
(\ref{wigner:master:squeezed}) is quite cumbersome, and we do not
explicitly write it here.
\par
Notice that if we suppose the environment composed by a set of
oscillators excited in a squeezed-thermal state of the form $\nu= S
(\zeta)\rho_{\rm th} S^\dag (\zeta)$, with $\zeta=|\zeta| e^{i\theta}$, 
$S(\zeta)=\exp\{\frac12  [\zeta^*\,a^{\dag
2}-\zeta\,a^2]\}$ and $\rho_{\rm th}=(1+n_{\rm th})^{-1} [n_{\rm th}/(1+n_{\rm
th})]^{a^\dag a}$, then we can rewrite the parameters $N$ and  $M$ in terms
of the squeezing and thermal number of photons $n_{\rm s}=\sinh^2 |\zeta|$ and
$n_{\rm th}$ respectively. We have $M=|M| e^{i\theta}$ \cite{grewal} and
\begin{eqnarray}
|M| =  \left(1 + 2\,n_{\rm th}\right)\sqrt{n_{\rm s}(1 +
n_{\rm s})}\,\, \hspace{0.5cm} \textrm{and} \hspace{0.5cm} \,\, N
= n_{\rm th} + n_{\rm s}(1 + 2\,n_{\rm th}) \:.
\end{eqnarray}
This parametrization automatically guarantees the semipositivity
of $\rho_t$.
%%%%%%%%%%%%%%%%%%%%%%%%%%%%%%%%%%%%%%%%%%%%%%%%%%%%%%%%%%%%%%%%%%%%%%%%%%
\section{Separability}\label{s:sep}
A quantum state of a bipartite system is {\em separable} if its
density operator can be written as $\varrho=\sum_k p_k \sigma_k
\otimes \tau_k$, where $\{p_k\}$ is a probability distribution and
$\tau$'s and $\sigma$'s are single-system density matrices. If a
state is separable the correlations between the two systems are of
purely classical origin, otherwise it is entangled. A
necessary and sufficient condition for separability of Gaussian
states is the positivity of the density matrix $\varrho^T$,
obtained by partial transposition of the original density matrix
(PPT condition) \cite{peres,geza,simon}. Notice that the Wigner 
function of a twin-beam is Gaussian and the evolution
in a Gaussian environment preserves such character. Therefore, we are
able to characterize the entanglement at any time and discuss its degradation
as a function of bath's parameters. The PPT condition for a density matrix can 
be rephrased as a condition on the covariance matrix $\mathbf{V}$ of the 
two-mode  Wigner function $W(x_1,y_1;x_2,y_2)$. After defining
\begin{eqnarray}
\begin{array}{cc}
\mathbf{\Omega} = \left( \begin{array}{cc} \mathbf{J} & \mathbf{0} \\
\mathbf{0} & \mathbf{-J}
\end{array} \right) &
\mbox{and} \quad
\mathbf{J} = \left( \begin{array}{cc} 0 & 1 \\
-1 & 0
\end{array} \right)\,.
\end{array}
\end{eqnarray}
and
\begin{eqnarray}
V_{pk} = \langle \Delta\xi_p\: \Delta\xi_k \rangle =
\int_{\mathbb{R}^4} {d}^4\xi\: \: \Delta\xi_p\: \Delta\xi_k\:
W(\xi)\,,
\end{eqnarray}
with $\Delta\xi_j = \xi_j - \langle \xi_j \rangle$, and
$\underline{\xi}=\{x_1,y_1,x_2,y_2 \}$, we have that a state is
separable iff
\begin{eqnarray} \label{15}
\mathbf{S}\equiv\mathbf{V}+\frac{i}{4}\, \mathbf{\Omega}\geq0
\end{eqnarray}
\par
\begin{figure}[h!]
\begin{center}
\includegraphics[width=0.7\textwidth]{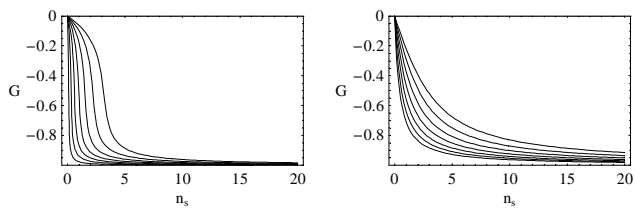}
\caption{Plots of the ratio $G = (t_{\rm s}-t_{0})/t_{0}$ as a
function of the number of squeezed photons $n_{\rm s}$ for
different values of the TWB parameter $\lambda$ and of the number
of thermal photons $n_{\rm th}$ when $M$ is real. The values of
$n_{\rm th}$ are $n_{\rm th} = 10^{-3}$ (left) and
$n_{\rm th}=1$ (right), while the solid lines, from bottom to top,
refer to $\lambda$ varying between 0.1 to 1.0 with steps of
0.15.}\label{f:threshold}
\end{center}
\end{figure}

For the state (\ref{wigner:evol:sq}), the condition (\ref{15}) 
rewrites as \cite{andre:sq}
\begin{eqnarray} \label{sigma:sep:cond}
\Sigma_{1}^{2}\:\Sigma_{4}^{2} \geq \frac{1}{16}\,,\quad\quad
\Sigma_{2}^{2}\:\Sigma_{3}^{2} \geq \frac{1}{16}\,.
\end{eqnarray}
We remind that in this case the squeezing parameter $M$ is real 
{\em i.e.} it has the same phase of the TWB parameter ($\theta =0$). By 
solving inequalities (\ref{sigma:sep:cond}) with respect to time 
$t$, we find that the TWB becomes separable for $t > t_{\rm s}$, 
where the survival time $t_{\rm s} = t_{\rm s}(\lambda,\Gamma,
n_{\rm th}, n_{\rm s})$ is given by
\begin{eqnarray}\label{tau:sep:squeezed}
t_{\rm s} = \frac{1}{\Gamma}\log\left( f + \frac{1}{1+2 n_{\rm
th}} \sqrt{f^2 + \frac{n_{\rm s}(1+n_{\rm s})} {n_{\rm
th}(1+n_{\rm th})}} \right)\,,
\end{eqnarray}
where we have defined
\begin{eqnarray}
f \equiv f(\lambda,n_{\rm th}, n_{\rm s}) =\frac{(1+2\,n_{\rm
th})\: \left[ 1+2\,n_{\rm th}-e^{-2\,\lambda}(1+2\,n_{\rm  s})
\right]} {4\,n_{\rm th}(1+n_{\rm th})}\,.
\end{eqnarray}
As one may expect, $t_{\rm s}$ decreases as $n_{\rm th}$ and
$n_{\rm s}$ increase.  Moreover, in the limit $n_{\rm\rm  s}
\rightarrow 0$, the threshold time reduces to the value $t_0$ 
pertaining a non squeezed bath \cite{andre:sq}. In order to see 
the effect of squeezing the bath on the survival time we introduce  
the function 
\begin{eqnarray} \label{26}
G(\lambda, n_{\rm th}, n_{\rm s}) &\equiv& \frac{t_{\rm s}-t_{\rm
0}}{t_0}\,.
\end{eqnarray}
$G>0$ means that squeezing leads to a longer survival time, shorter 
otherwise. Results are illustrated in Fig. \ref{f:threshold}, where 
we plot $G$ as a function of $n_{\rm s}$ for different values of $n_{\rm
th}$ and $\lambda$. Since $G$ is always negative, we conclude that
coupling a TWB with an {\em in-phase} squeezed bath destroys 
entanglement faster than the coupling with 
a non squeezed environment.
\par
\begin{figure}[h!]
\begin{center}
\includegraphics[width=0.5\textwidth]{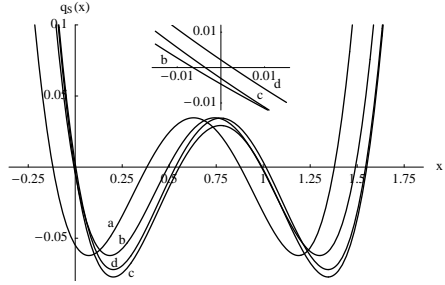}
\caption{Plots of the characteristic polynomial
$q_{\mathbf{S}}(x)$ associated to $\mathbf{S}$ for a TWB
propagating in a non classical environment with squeezing
parameter $M = |M|\, e^{i\theta}$. We set $e^{-\Gamma t} = 0.55$,
$\lambda=1$, and (a) $n_{\rm th}=n_{\rm s}=0$, (b) $n_{\rm
th}=0.5$, $n_{\rm s}=0$, (c) $n_{\rm th}=0.5$, $n_{\rm s}=0.07$
and $\theta = 0$, (d) $n_{\rm th}=0.5$, $n_{\rm s}=0.07$ and
$\theta = \pi/5$. Notice that there are always four real roots and
three of them are positive. The inset is a magnification of
the region near to 0: the presence of thermal noise and non
classical fluctuations with non zero phase reduces the survival
time.} \label{f:theta}
\end{center}
\end{figure}
\begin{figure}[h!]
\begin{center}
\includegraphics[width=0.5\textwidth]{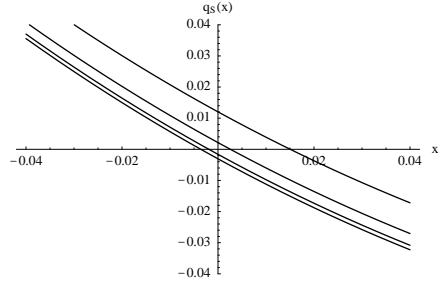}
\caption{Plots of the characteristic polynomial
$q_{\mathbf{S}}(x)$ associated to $\mathbf{S}$ for a TWB
propagating in a non classical environment with squeezing
parameter $M = |M|\, e^{i\theta}$. 
This plot only shows the region
near to $x=0$ (the other three roots are always positive). We set
$e^{-\Gamma t} = 0.55$, $\lambda=1$, $n_{\rm th}=0.5$, $n_{\rm
s}=0.07$ and, from left to right, $\theta = 0, \pi/10, \pi/5$ and
$\pi/2$: a non real squeezing parameter of the bath always reduces
the survival time. \label{f:var:theta}}
\end{center}
\end{figure}

We have also evaluated the threshold time for separability in 
case of an {\em out-of-phase} squeezed bath {\em i.e.} 
for complex $M=|M|\,e^{i \theta}$.
The analytical expression is quite cumbersome and will not 
be reported here. However, in order to investigate the positivity 
of $\mathbf{S}$ as a function of $\theta$, it suffices to consider   
the characteristic polynomial $q_{\mathbf{S}}(x)$ associated to
$\mathbf{S}$, and study the sign of its roots. A numerical
analysis shows that this polynomial has four {\em real} roots and
three of them are always {\em positive}. We focus our attention on
the other one. In Fig. \ref{f:theta} we plot the
characteristic polynomial for different values of the parameters
$n_{\rm th}$, $n_{\rm s}$ and $\theta$ (we put $e^{-\Gamma t} =
0.55$): it is apparent that by adding thermal
noise and non classical fluctuations
the sign of the smallest root changes from negative (entangled
state) to positive (separable state). In
other words, the survival time becomes shorter. Moreover, in
Fig. \ref{f:var:theta} we show that increasing $\theta$ from $0$
to $\pi/2$ the threshold time is further reduced. The behavior 
of the polynomial roots for different values of $\lambda$ and $\Gamma t$
is analogue.
%%%%%%%%%%%%%%%%%%%%%%%%%%%%%%%%%%%%%%%%%%%%%%%%%%%%%%%%%%%%%%%%%%%%%
\section{Conclusions}\label{s:conclusions}
In this paper we have analyzed the propagation of a TWB
through Gaussian phase-sensitive noisy channels, and have 
evaluated the threshold (survival) time for the state to become 
separable.
\par
We found that the survival time in a squeezed environment  
is always shorter than in a purely dissipative or thermal ones.
In addition, the survival time is further reduced 
if the squeezing phase of the fluctuations is different from the TWB 
phase.
%%%%%%%%%%%%%%%%%%%%%%%%%%%%%%%%%%%%%%%%%%%%%%%%%%%%%%%%%%%%%%%%%%%%

%%%%%%%%%%%%%%%%%%%%%%%%%%%%%%%%%%%%%%%%%%%%%%%%%%%%%%%%%%%%%%%%%%%%%%
%%%%%%%%%%%%%%%%%%%%%%%%%%%%%%%%%%%%%%%%%%%%%%%%%%%%%%%%%%%%%%%%%%%%%%%
%%%%%%%%%%%%%%%%%%%%%%%%%%%%%%%%%%%%%%%%%%%%%%%%%%%%%%%%%%%%%%%%%%%%%%%
%%%%%%%%%%%%%%%%%%%%%%%%%%%%%%%%%%%%%%%%%%%%%%%%%%%%%%%%%%%%%%%%%%%%%%%
\end{document}